\documentclass[doublecol,citesort]{epl2} 

\usepackage{graphicx}
\usepackage{CJK} 
\usepackage{amsmath}
 \usepackage{amssymb}
\usepackage{epstopdf}
\usepackage{color}
\usepackage{xcolor}
\usepackage{float}
\usepackage{bm}
\usepackage[normalem]{ulem}

\usepackage{hyperref}

\title{Equilibrium states of the ice-water front\\ in a differentially heated rectangular cell}	

\shorttitle{}
		
\author{Ziqi Wang\inst{1} 
	\and Enrico Calzavarini\inst{2} \footnote{enrico.calzavarini@univ-lille.fr}
			\and Chao Sun\inst{1,3}\footnote{chaosun@tsinghua.edu.cn}
		}
	
\shortauthor{Z. Wang \etal}

\institute{                    
  \inst{1} Center for Combustion Energy, Key laboratory for Thermal Science and Power Engineering of Ministry of Education, Department of Energy and Power Engineering, Tsinghua University, Beijing 100084, China\\
  \inst{2} Univ. Lille, Unit\'e de M\'ecanique de Lille  J. Boussinesq - UML - ULR 7512, F-59000 Lille, France\\ 
  \inst{3} Department of Engineering Mechanics, School of Aerospace Engineering, Tsinghua University, Beijing 100084, China
}
\pacs{nn.mm.xx}{Rayleigh-B\'enard convection, Multiphase flows, Freezing, Melting}

\abstract{We study the conductive and convective states of phase-change of pure water in a rectangular container where two opposite walls are kept respectively at temperatures below and above the freezing point and all the other boundaries are thermally insulating.
	The global ice content at the equilibrium and the corresponding shape of the ice-water interface are examined, extending the available experimental measurements and numerical simulations. 
	We first address the effect of the initial condition, either fully liquid or fully frozen, on the system evolution. Secondly, we explore the influence of the aspect ratio of the cell, both in the configurations where the background thermal-gradient is antiparallel to the gravity, namely the Rayleigh-B\'enard (RB) setting, and when they are perpendicular, i.e., vertical convection (VC).
	We find that for a set of well-identified conditions the system in the RB configuration displays multiple equilibrium states, either conductive rather than convective, or convective but with different ice front patterns. 
	The shape of the ice front appears to be always determined by the large scale circulation in the system. In RB, the precise shape depends on the degree of lateral confinement. 
	In the VC case the ice front morphology is more robust, due to the presence of two vertically stacked counter-rotating convective rolls for all the studied cell aspect-ratios.
 }

\begin{document}
	\maketitle

	\textbf{Introduction.} -- 
Convective liquids undergoing melting or freezing give rise to a rich phenomenology of flow patterns and solid-phase morphologies.
This bears great relevance in the geophysical domain, e.g., in volcanology for the understanding of the solidification of magma chambers \cite{brandeis1986interaction,brandeis1989convective,sparks2019formation}, in planetology for the study of magma oceans \cite{ulvrova2012numerical}, or in geomorphology and glaciology for glacier dynamics and their induced erosion \cite{meakin2010geological}, and in marine sciences for the prediction of arctic sea-ice annual cycles \cite{deser2000arctic}. In the technological context, convection driven phase-change has a key role in metallurgy \cite{worster1997convection}, in the purification of substances \cite{glicksman2010principles} and in the storage of thermal energy through phase-change materials \cite{nazir2019recent}.
Depending on the application, the fluid can be either very complex in composition and rheology like a heterogenous molten rock magma, or simpler like muddy-water or sea-water, or nearly ideal as for purified materials employed in chemical industrial processes. 
However, probably the most common instance is the one where the fluid is relatively clean water. This is a rather intriguing case due to the peculiar non-monotonous buoyancy force intensity in water above the freezing point.
A large variety of situations of thermal or mechanical driving mechanisms are encountered. The fluid motion can be the result of natural convection due to a cooling process, as in the case of volcanic magma, or it can be steadily driven by localized heat sources, such as a hot/cold boundary, or by distributed ones, as in the case of internal heating by radioactive decay (common in rock formation) or by absorption of solar radiation as it happens for water in glaciers or in the oceans \cite{skyllingstad2007numerical,scagliarini2020modelling}.
Finally, the thermal convection can also be maintained by a mechanical driving of the fluid, as it happens below the sea-ice and around icebergs \cite{hester2021aspect}.
From a physicist's perspective, a relevant question is how the convecting fluid flow, which can be either laminar or turbulent depending on the driving intensity,  can determine the overall shape of ice. 
Is there just a single general mechanism? Can one envisage a phenomenological model, i.e. a predictive argument for the shape of the ice-water front that does not need to take into account the exceptional complexity of the thermodynamics and fluid dynamics equations involved in its description? 

A way to reply to these questions is to envisage a sufficiently simplified and well-controllable system that retains part of the physical complexity and rich phenomenology observed in nature. A well studied and commonly used convective setup in fluid-dynamics is the  
Rayleigh-B\'enrad (RB) system, a fluid filled container heated at the bottom surface (of temperature $T_b$) and cooled at the top (of temperature $T_t$) and thermally insulated on the lateral sides.
The thermal instability, the onset of convection, the flow bifurcations, the self-organization of the system in distinct parts, the turbulent regime and its scaling laws, have been analyzed in great detail \cite{sig94,bod00,ahlers2009heat,lohse2010small,chi12,kadanoff2001turbulent}.
This makes RB an ideal system on top of which the complexity of phase-change phenomena and its coupling with fluid convection can be added to. Previous studies of melting in RB have highlighted the complexity of the solid-liquid interface topography, and its dependencies on the strength of convection (parametrized by the Rayleigh number), on material properties (Stefan and Prandtl numbers), on the system dimensionality and geometry\cite{Esfahani2018Basal,favier2019rayleigh,satbhai2019comparison,2020_Bistability,toppaladoddi_2021}. Furthermore, an intriguing phenomenon of bistability depending on whether the system has been initialized as fully liquid or fully frozen has been identified \cite{vasil2011dynamic, 2020_Bistability}.
However, most of the above mentioned studies have focused on the case of pure materials and small thermal gaps so that the resulting buoyancy is linear with respect to temperature (Boussinesq approximation). For water, the density anomaly, i.e. the local mass density maximum at around $T_c \sim 4^\circ$C at atmospheric pressure, leads to a non-linear buoyancy force, which plays a great role in the icing dynamics. Recently, \cite{wang2020ice} combined experiments and numerical simulations to understand the hydrodynamical mechanisms that control the global extent and the shape of the ice in a freezing RB rectangular quasi-two-dimensional cell. They identified four distinct regimes of coupling between thermal stratification, buoyancy forces, and phase-change \cite{wang2020ice} (see Table.~\ref{regimes} for a brief summary). 
Other research efforts have explored the effect of the RB system inclination on the overall icing process \cite{wang2021growth}. 

In this work, we aim at extending further the available experimental measurements and numerical simulations of water melting and icing in an RB cell, in particular by testing the existence of the bistability in a water-filled RB cell and at exploring the container aspect ratio dependence and its role with respect to the system inclination, either straight or rotated by 90$^\circ$.

\begin{table*}[htbp]
	\begin{center}
		{\small
			
			\begin{tabular}{| c | c | c | c | c |}
				\hline
				Regime & 1 & 2 & 3 & 4\\
				\hline
				Stratification type & SSD & SSD(top) + USD(bottom)& SSD(top) + USC(bottom)& SSD(top) +USC(bottom)\\
				\hline
				Equilibrium state & Diffusive & Diffusive & Convective & Convective \\
				\hline
				Ice front shape& Flat & Flat & Flat & Deformed\\
				\hline	
			\end{tabular}
		}
		\caption{Four regimes of water-freezing RB system for increasing temperature differences (left to right) as identified in \cite{wang2021growth}: in the second line, the first two letters of the acronyms specify the stratification type, either stably stratified (SS, with temperature ranging from $T_0$ to $T_c$) or unstably stratified (US, with temperature ranging from $T_c$ to $T_b$); the third letter specifies the mode of heat transport
			, either diffusion (D) or convection (C); top and bottom denote the positions of the layers in the RB system.
		}
		\label{regimes}
	\end{center}
\end{table*}

\begin{figure*}[t!hb]
	\centering
	\includegraphics[width=1\textwidth]{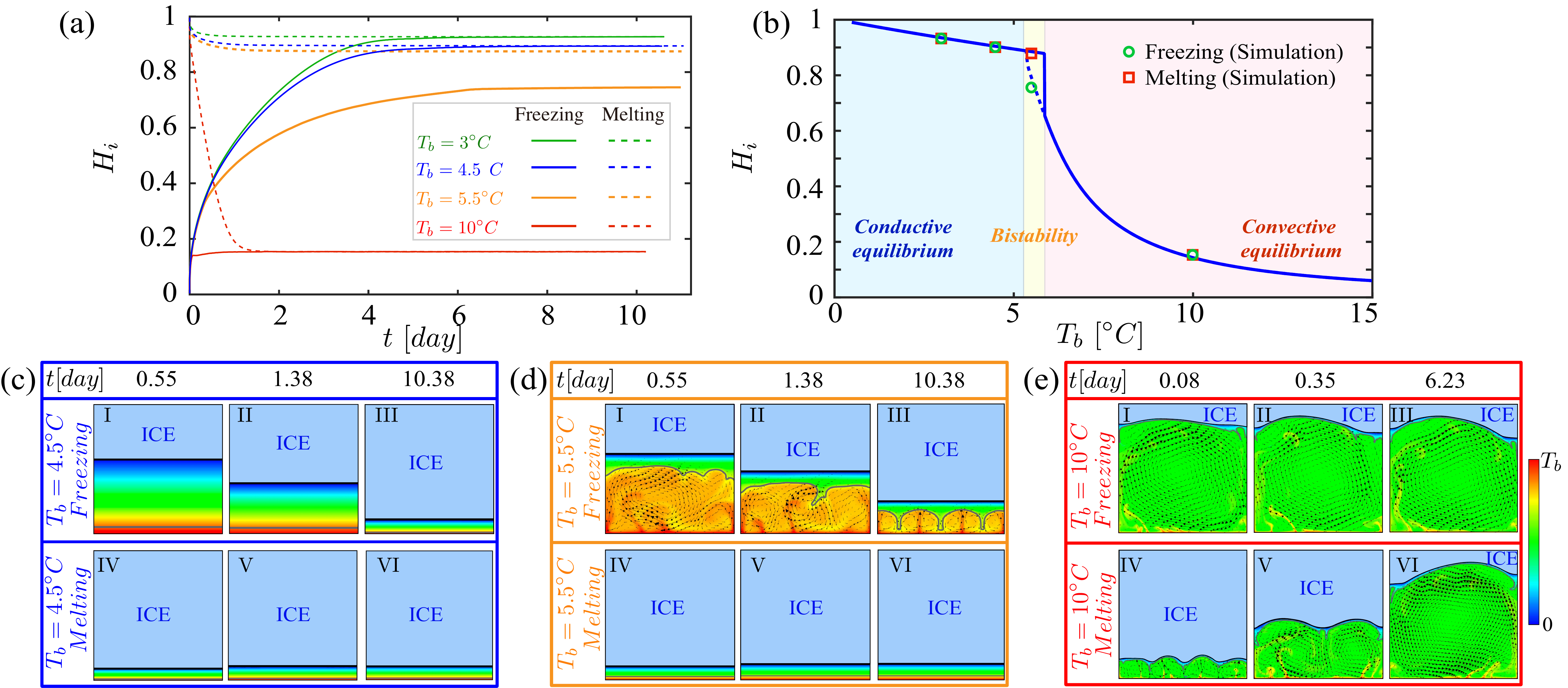}
	\caption{Bistability phenomenon. (a) Temporal evolution of the ice thickness, $H_i$, of the freezing process (thick line) and the melting process (dashed line). 
		(b) Comparison of the simulation (symbols) and the theoretical model (dashed and solid lines):
		$H_i$ as a function of $T_b$. The blue, yellow, and red shaded areas correspond to the conductive, bisable and convective equilibrium regimes, respectively. The dashed line denotes one of the two branches in the bistable regime which predicts the convective equilibrium state.			
		(c), (d) and (e): panels \text{\uppercase\expandafter{\romannumeral1}}-\text{\uppercase\expandafter{\romannumeral3}} display the evolution of the instantaneous temperature field of the freezing process, with the bottom plate temperature $T_b =  4.5, 5.5, 10^\circ$C, respectively; panels \text{\uppercase\expandafter{\romannumeral4}}-\text{\uppercase\expandafter{\romannumeral6}} display the evolution of the instantaneous temperature field of the melting process, for the same bottom plate temperatures.
		The visualizations of the temperature field in the water region are with colors and $T_0$ (black line) and $T_c$ (gray line) isotherms, along with the velocity vectors when fluid convection occurs. 
	}
	\label{fig2} 
\end{figure*}	

\begin{figure}[t!hb]
	\centering
	\includegraphics[width=1\linewidth]{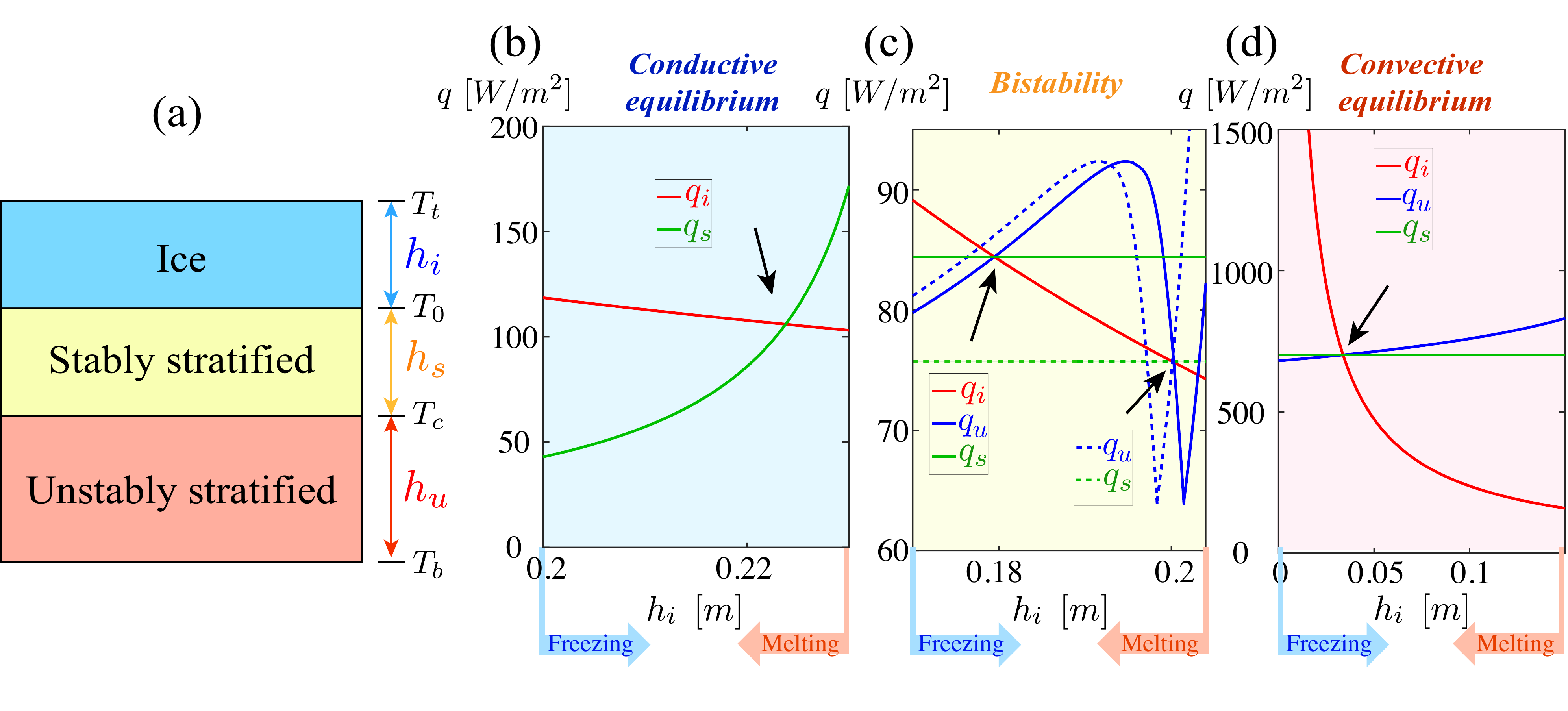}
	\caption{
		Mechanism of bistability: (a) Sketch of the one-dimensional model: three plane layers are connected in series. Note that if  $T_b < T_c$ the unstably stratified layer is absent; The computed heat-fluxes across the cell as a function of the ice thickness, $h_i$, in the purely conductive regime (b), the bistable regime (c) and the convective one (d).
	}
	\label{figadd} 
\end{figure}	

\begin{figure*}[t!hb]
	\begin{center}
		\includegraphics[width=1.0\textwidth]{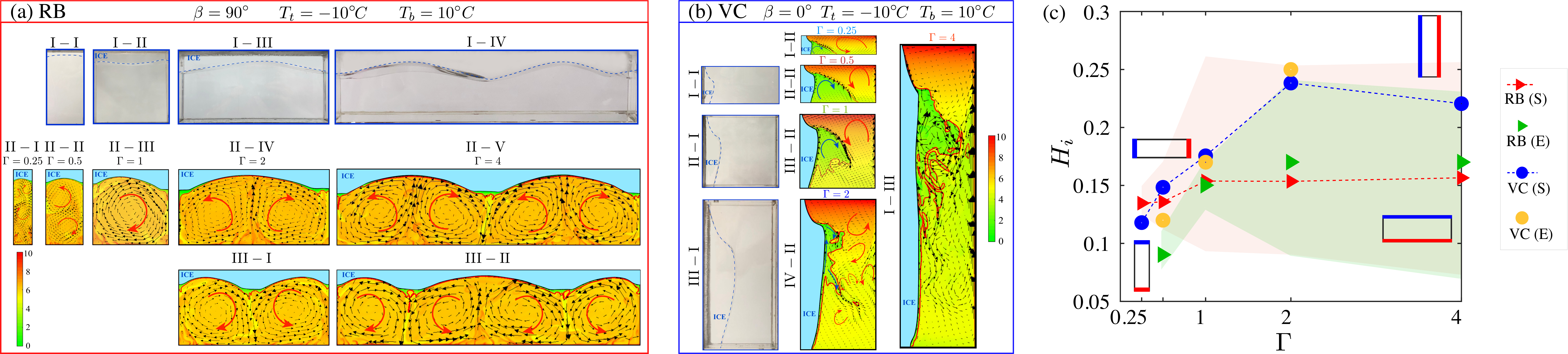}
		\caption{Aspect ratio dependence. (a) The temperature field at the equilibrium state of the RB case.	
			Panels \text{\uppercase\expandafter{\romannumeral1}}-\text{\uppercase\expandafter{\romannumeral1}},  \text{\uppercase\expandafter{\romannumeral1}}-\text{\uppercase\expandafter{\romannumeral2}},  \text{\uppercase\expandafter{\romannumeral1}}-\text{\uppercase\expandafter{\romannumeral3}}, and \text{\uppercase\expandafter{\romannumeral1}}-\text{\uppercase\expandafter{\romannumeral4}} display the experimental results with $\Gamma = 0.5, 1, 2,4$, respectively. Panels \text{\uppercase\expandafter{\romannumeral2}}-\text{\uppercase\expandafter{\romannumeral1}},  \text{\uppercase\expandafter{\romannumeral2}}-\text{\uppercase\expandafter{\romannumeral2}}, \text{\uppercase\expandafter{\romannumeral2}}-\text{\uppercase\expandafter{\romannumeral3}}, \text{\uppercase\expandafter{\romannumeral2}}-\text{\uppercase\expandafter{\romannumeral4}}, and \text{\uppercase\expandafter{\romannumeral2}}-\text{\uppercase\expandafter{\romannumeral5}} display the simulation results,
			with $\Gamma = 0.25, 0.5, 1, 2,$ and 4, respectively. Panels \text{\uppercase\expandafter{\romannumeral3}}-\text{\uppercase\expandafter{\romannumeral1}} and \text{\uppercase\expandafter{\romannumeral3}}-\text{\uppercase\expandafter{\romannumeral2}}
			displays the simulation results for $\Gamma = 2$ and 4, same conditions as \text{\uppercase\expandafter{\romannumeral2}}-\text{\uppercase\expandafter{\romannumeral4}} and \text{\uppercase\expandafter{\romannumeral2}}-\text{\uppercase\expandafter{\romannumeral5}} but with different ice front shapes.
			(b) The temperature field at the equilibrium state of the VC case.
			Panels \text{\uppercase\expandafter{\romannumeral1}}-\text{\uppercase\expandafter{\romannumeral1}},  \text{\uppercase\expandafter{\romannumeral2}}-\text{\uppercase\expandafter{\romannumeral1}}, and \text{\uppercase\expandafter{\romannumeral3}}-\text{\uppercase\expandafter{\romannumeral1}} display the experimental results with $\Gamma = 0.5, 1, 2$, respectively. Panels \text{\uppercase\expandafter{\romannumeral1}}-\text{\uppercase\expandafter{\romannumeral2}}, 
			\text{\uppercase\expandafter{\romannumeral2}}-\text{\uppercase\expandafter{\romannumeral2}},  \text{\uppercase\expandafter{\romannumeral3}}-\text{\uppercase\expandafter{\romannumeral2}},  \text{\uppercase\expandafter{\romannumeral4}}-\text{\uppercase\expandafter{\romannumeral2}}, and \text{\uppercase\expandafter{\romannumeral1}}-\text{\uppercase\expandafter{\romannumeral3}} display the simulation results, with $\Gamma = 0.25, 0.5, 1, 2,$ and 4, respectively.
			The visualizations of the temperature field in (a-b) display $T_0$ (black line) and $T_c$ (red line) isotherms.
			(c) $H_i$ as a function of $\Gamma$ for RB and VC cases from simulations (S) and experiments (E).
			The shaded area shows the spatial variation of the ice front (from minima to maxima of $h_i(x,t\to \infty)/h$) in the RB case for experiments (green) and simulations (red).  
		}
		\label{fig3} 
	\end{center}
\end{figure*}	

\begin{figure}[t!hb]
	\centering
	\includegraphics[width=1\linewidth]{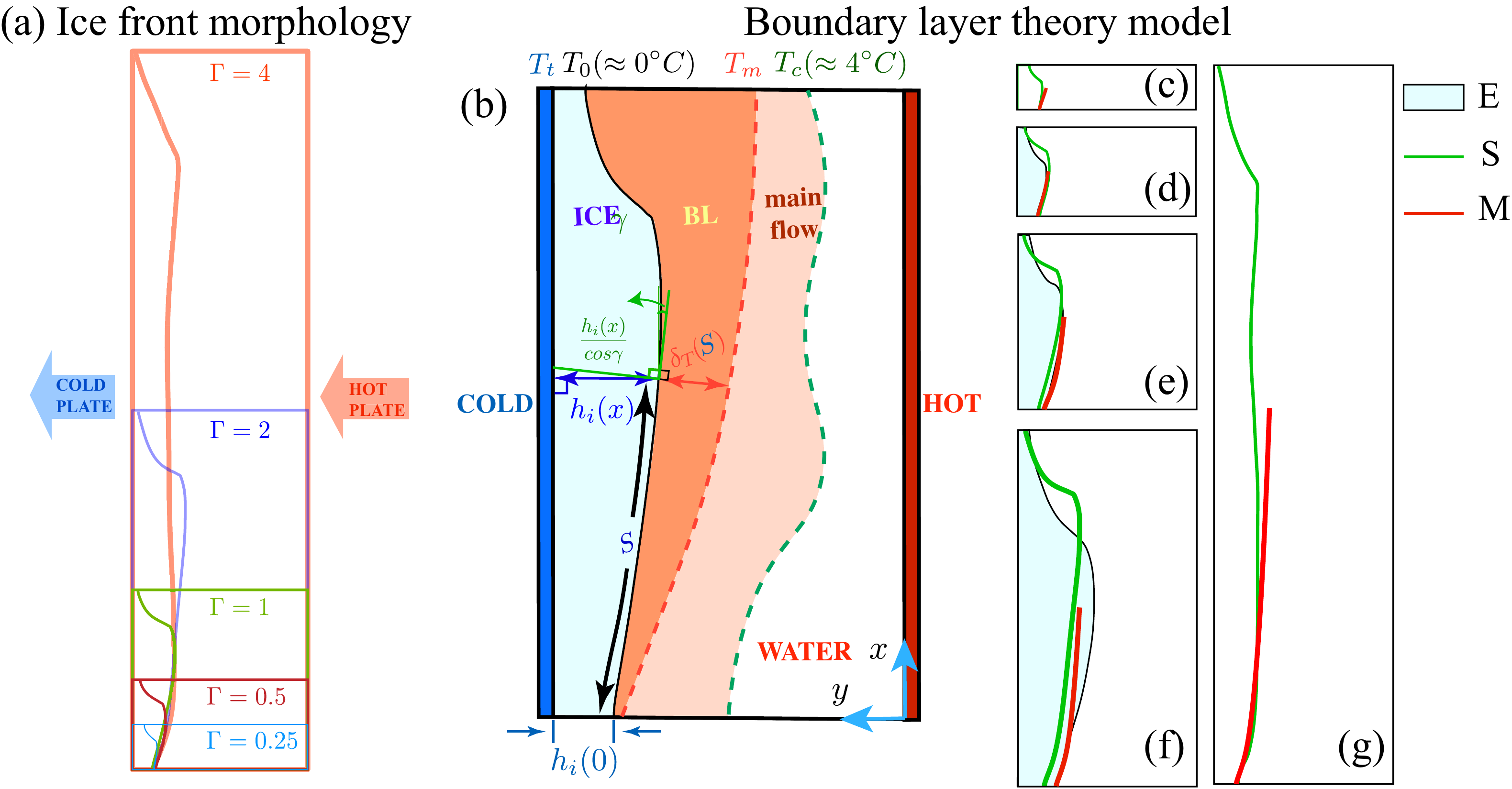}
	\caption{Ice front shape boundary layer based model. (a) Ice front morphology at the statistical equilibrium in VC as extracted from the numerical simulation with aspect ratio, $\Gamma=$ 0.25 (light blue line), 0.5 (red line), 1 (green line), 2 (dark blue line), 4 (orange line). The system is cooled at the left side and heated at the right side. 
		(b) Sketch of the phenomenological BL model:
		the ice front at $T_0 = 0^\circ$C (black line), 
		{\color{black}the $T_m$ isotherm which is the outer boundary of the thermal boundary (red dashed line),}
		the $T_c$ isotherm (green dashed line). The angle between the tangential direction of the ice front and the x-direction is $\gamma$; the thickness normal to the ice front (the green thick line in the ice) is $\frac{h_i(x)}{cos\gamma}$ (with $cos\gamma = \frac{dS}{dx}$).
		(c)-(g) Comparison of ice front morphology between experiments (blue shaded area), simulations (green line) and the model (red line) for $\Gamma= 0.25$ (c); 0.5 (d); 1 (e); 2 (f); 4(g). 
	}
	\label{fig4} 
\end{figure}

\textbf{Experimental setup and Numerical Methods.} -- 
The experiments are conducted in a battery of standard RB cells with four different geometrical proportions of the container. 
The details of the setup are documented in the Supplementary Information (SI). Here we only give a brief description.
The cell aspect ratio, $\Gamma=l_x/h$, is here defined as the quotient between the largest width of the isothermal plates ($l_x$) and the distance between the isothermal plates (also called height) $h$. We choose $l_x = 24$ cm and $h = 12, 24, 48, 96$ cm leading to $\Gamma=0.5, 1, 2, 4$ respectively. The third spatial dimension, $l_z = 6$ cm, is kept fixed and it is smaller than the others, a fact that allows to designating the system as quasi-two-dimensional. We note that $h$ stands for the vertical direction, parallel to the gravitational acceleration, in the RB configuration, while it is horizontal in the vertical convection (VC) setting, when the same experimental cells are tilted by $90$ degrees. 
In the present study, we only perform solidification experiments, i.e. experiments whose initial condition is liquid water with uniform initial temperature same as the hot plate temperature $T_b$. 
Given the difficulty of manufacturing a cell-sized homogeneous ice block, we could not perform (well controlled) melting experiments.  
For this reason, in the present study we also rely on the results of numerical simulations. We employ the \textsc{CH4-project} code \cite{calzavarini2019eulerian} which is based on a Lattice-Boltzmann equation algorithm for the description of fluid and temperature dynamics, and on the enthalpy method for phase-change. Both the numerical methods and the specific algorithm implementation have been extensively validated, particularly against RB water freezing experiments in \cite{wang2021growth, wang2020ice}.
We use the non-monotonic relationship of the water density as a function of the temperature, $\rho = \rho_c(1-\alpha^*|T_b-T_c|^q )$, 
with $\rho_c = 999.972~kg/m^3$ the maximum density corresponding to $T_c \approx 4^\circ$C, $\alpha^* = 9.30\times10^{-6} (K^{-q})$, and $q=1.895$ \cite{Gebhart1977A}. 
We conduct two-dimensional simulations, which have been proved to match well the experiments for the present quasi-2D cells \cite{wang2021growth, wang2020ice}. The initial conditions encompass both the fully liquid and fully frozen cases, i.e., freezing and melting numerical experiments.
Throughout the investigation, the cold plate is fixed both in the experiments and the simulations at temperature $T_t = -10^\circ$C. The effects of changing $T_t$ are qualitatively predictable (more details on $T_t$ effects can be found in SI).
Both in the experiments and in the simulations we monitor the evolution of the ice-water interface $h_i(x,t)$ ($i$ stand for ice), and its mean instantaneous value
profile $\overline{h_i}(t) = l_x^{-1}\int_0^{l_x} h_i(x,t) dx$. Such quantities are more conveniently analyzed in dimensionless terms, hence $H_i = \overline{h_i}/h$ represents the fraction of the cell that is occupied by the ice 
($H_i =1$ means complete solidification). 

\textbf{Results: bistability.} -- We first study whether the melting and freezing system configurations evolve into the same equilibrium state.
Given the fact that the system in the RB arrangement can display different regimes as the intensity of the thermal driving increases
(Table.~\ref{regimes}), we choose four typical cases of the thermal driving condition from each of the existing four different regimes, i.e., $T_b = 3, 4.5, 5.5, 10^\circ$C, and perform long-term numerical simulations for melting (initial condition is $H_i=1$) and freezing (initial condition is $H_i=0$). 
Note that the duration of the simulations is of the order of 10 days in physical units.
Fig.~\ref{fig2}(a) shows the normalized global ice thickness, $H_{i}$, as a function of time under different conditions of thermal driving in the RB setup. The thick (dashed) lines represent the freezing (melting) case. It is observed that melting and freezing reach the same equilibrium state for the cases of $T_b = 3, 4.5, 10^\circ$C; while for $T_b =5.5^\circ$C, melting and freezing reach a conductive and a convective equilibrium respectively. Hence, in the freezing experiments the global amount of ice at the equilibrium is smaller.
To better understand this phenomenon, we show the evolution of the temperature field (see Fig.~\ref{fig2}(c)-(e), we do not show the $T_b=3^\circ$C case because when $T_b< T_c$ (Regime-1), 
the fluid region is trivially stably-stratified  throughout the entire simulation). 
At $T_b=4.5^\circ$C (Regime-2), though the stably and unstably-stratified layers coexist, the effective Rayleigh number (which is based on the unstably-stratified layer and only defined when $T_b > T_c$), $Ra_e = \frac{(\Delta\rho/\rho_\text{c})g(h_c)^3}{\nu \kappa}=\frac{g\alpha^* |T_\text{b}-T_\text{c}|^q (h_c)^3 }{\nu \kappa}$ (with $h_c$ being the averaged height of the $T_c$ isotherm, $g$ the gravitational acceleration, $\nu$ the kinematic viscosity, and $\kappa$ the thermal diffusivity), decreases from $Ra_e\sim10^3$ as the ice thickness increases in the freezing case (Fig.~\ref{fig2}(c)Panels \text{\uppercase\expandafter{\romannumeral1}}-\text{\uppercase\expandafter{\romannumeral3}}), while increases to $Ra_e\sim 10$ as the ice thickness decreases in the melting case (Fig.~\ref{fig2}(c)Panels \text{\uppercase\expandafter{\romannumeral4}}-\text{\uppercase\expandafter{\romannumeral6}}). Throughout the process of melting and freezing, the thermal driving remains below the critical Rayleigh number ($Ra_{cr}$ is $\sim 2585$ in a laterally closed unit aspect ratio cell \cite{luijkx1981onset},
$\sim1708$ in the laterally infinite RB system \cite{chandrasekhar2013hydrodynamic}, 
and $\sim 1493$ in a laterally infinite RB with linear buoyancy  and melting \cite{2020_Bistability}),
and thus conduction holds \cite{satbhai2020criteria}. 
Similarly, when the thermal driving is strong , i.e., $T_b=10^\circ$C, (Regime-4, see Fig.~\ref{fig2}(e)),  the estimated effective Rayleigh number is $Ra_e\sim 10^7\gg Ra_{cr}$ at the final equilibrium state, the turbulent developed convection state prevails both for the freezing and melting numerical experiments.
However, freezing and melting behave differently in Fig.~\ref{fig2}(d) (Regime-3). For the freezing case, as the ice grows, the effective aspect ratio (i.e., $l_x/(h-h_i)$) 
of the fluid region increases, and thus the convective rolls self-organize into several smaller rolls and the $Ra_e$ decreases accordingly, but the $Ra_e$ is still $\sim 4\times 10^3 > Ra_{cr}$, leading to retaining convection. While for the melting case, the system reaches an equilibrium when it is still in the conductive state. 

Additionally, we explore the VC configuration in the same conditions. Differently from the RB case, the global ice thickness $H_i$ always reaches the same final equilibrium independently of the initial conditions or independently of the history of the system ({\color{black}see SI for the corresponding visualizations}). In VC, the temperature gradient is perpendicular to the gravity, making the system thermally unstable for any temperature gap or $Ra$. This leads to the immediate breakdown of the stable stratification that is observed in the RB. 

The bistability observed in the phase-change RB can be quantitatively predicted by means of a one-dimensional heat flux model. This is as follows: We consider the system as the juxtaposition in series of compartments each one characterized by specific thermal properties (we neglect the curvature of the ice-water interface and the interface between the stably- and unstably-stratified layers). 
The system reaches an equilibrium state when there is heat flux balance among the conductive heat flux through the ice layer (denoted $q_i$), the conductive heat flux through the stably stratified liquid layer ($q_s$), and the either conductive or convective heat flux (depending on $Ra_e$) of the unstably stratified layer ($q_u$). Note that  $q_u$  exists only when $T_b>T_c$. The corresponding thicknesses for each layer are denoted as $h_{i}$, $h_s$, and $h_u$ (see Fig.~\ref{figadd}(a)). Depending on whether $T_b$ is above $T_c$, two situations need to be considered: 1) $T_b \leq T_c$, the fluid layer is gravitationally stable, so the heat transfers diffusively both in the ice and water, so the definitions of the heat fluxes are $q_i = k_{i} \frac{T_0 - T_t}{h_i}$ and $q_s =  k_{w} \frac{T_b -T_0}{h_\text{s}}$, where $k_{i}$ and $k_{w}$ are the thermal conductivity of ice and water, respectively; 2) $T_b > T_c$, the density gradient takes opposite signs in the layers of temperature ranging from $T_0$ to $T_c$ (gravitationally stably-stratified) and that ranging from $T_c$ to $T_b$ (gravitationally unstably-stratified). 
The heat flux in the gravitationally unstably-stratified layer can be predicted based on the relation of the effective Nusselt number, $Nu_e$ (the dimensionless heat flux normalized by the diffusive heat flux based on the unstably-stratified layer), and the $Ra_e$. So the definitions of the heat fluxes are $q_i = k_{i} \frac{T_0 - T_t}{h_i}$, $q_s =  k_{w} \frac{T_c -T_0}{h_\text{s}}$, and $q_u = \text{Nu}_e k_{w}  \frac{T_\text{b}-T_\text{c}}{h_\text{u}}$, where $Nu_e$ as a function of the $Ra_e$ behaves as in the classical RB convection  (as we already studied in \cite{wang2021growth}). 
In summary, depending on the value of $T_b$ the equilibrium condition reads:

\textit{Case-1: $T_b \leq T_\text{c}$}.
The system is in a conductive state and is independent of the water layer thickness, the heat flux balance reads
\begin{equation}
	q_i =  q_s.
	\label{eqdiffusive}
\end{equation}
From Eq.~(\ref{eqdiffusive}) we obtain $h_i=  \frac{-k_{i} T_t}{k_{w} T_b - k_{i} T_t} h$.

\textit{Case-2: $T_b > T_\text{c}$}. 
\begin{equation}
	q_i = q_s=q_u.
	\label{eqconvective}
\end{equation}
From Eq.~(\ref{eqdiffusive}) and Eq.~(\ref{eqconvective}), $h_i$ can be calculated which is reported in Fig.~\ref{fig2}(b). The theoretical prediction (dashed and thick lines) agrees well with the numerical simulation results (symbols). As $T_b$ increases, the system experiences three regimes: 1) the equilibrium state is conduction independently of the freezing or melting initial system configuration (blue shaded area in Fig.~\ref{fig2}(b)); 2) the freezing case reaches the convective equilibrium state (dashed line in Fig.~\ref{fig2}(b)), while the melting case reaches the conductive equilibrium state, and this phenomenon is denoted by the bistability (yellow shaded area in Fig.~\ref{fig2}(b)); 3) the convective equilibrium is reached independently of freezing or melting (red shaded area in Fig.~\ref{fig2}(b)). This bistable regime which has been recently observed in numerical simulation of fluids with a linear buoyancy force \cite{2020_Bistability}, is here verified in the case of freezing pure water.

The one-dimensional model also allows us to understand which equilibrium state is reached depending on the initial conditions. This is illustrated in Fig.~\ref{figadd}, where we plot the values of the heat fluxes across the layers, $q_{i}$, $q_{s}$, $q_{u}$ as a function of the ice thickness $h_i$.
We note that in a freezing experiment $h_i$  is initially null and shall grow till a matching of the fluxes is reached, and vice-versa, i.e., in melting experiment $h_i = h$ and decreases over time till the flux matching occurs. While the matching is unique in the Case-1 (Fig.~\ref{figadd}(b)) and for strongest convection (Fig.~\ref{figadd}(d)),  the freezing front (bottom blue arrow) encounters first an equilibrium position with smaller $h_i$ than the melting case (bottom red arrow) (Fig.~\ref{figadd}(c)) in the  Case-2 with moderate thermal forcing.

\textbf{Results: aspect ratio dependence.} -- We now focus on the case freezing of an initially liquid-filled cavity with large thermal driving (high $Ra$). We
perform both experiments and numerical simulations of the aspect ratio dependence in the RB and VC arrangements.
The overall maximal variations of $H_i$ for different $\Gamma$ are modest ($< 20\%$) for the RB case, while they can be large (up to $200 \%$) for the VC case.
For the RB, it is observed that as $\Gamma \geq 1$, the global ice content nearly saturates (Fig.~\ref{fig3}(c)), and the only effect induced by larger $\Gamma$ is that the number of the corresponding large-circulation convective rolls increases, with each roll of about unit width-to-height ratio (see the one, two, and four convective rolls for $\Gamma =1, 2, 4$, respectively, in the visualizations of the temperature field from simulations in Fig.~\ref{fig3}(a) panels \text{\uppercase\expandafter{\romannumeral2}}-\text{\uppercase\expandafter{\romannumeral3}},  \text{\uppercase\expandafter{\romannumeral2}}-\text{\uppercase\expandafter{\romannumeral4}}, and \text{\uppercase\expandafter{\romannumeral2}}-\text{\uppercase\expandafter{\romannumeral5}}). 
We also remark that the ice is locally flat in between the neighboring convective rolls, in correspondence to the position from which cold plumes detach. 
The characteristics of the ice front morphology are due to the shield of the stably-stratified layer (with temperature ranging from $T_0$ to  $T_c$). Firstly, acting as a buffer layer, the stably stratified layer is able to alleviate the effect of the hot plume impingement, creating a locally cold environment; 
secondly, at the edge of the two neighboring large scale rolls, the penetration from the unstably stratified layer does not occur, 
and this creates a locally quiescent region, contributing to the local flattening of the ice (however evident only in the simulations).
Another interesting feature is that multiple forms of stable ice profiles can be observed corresponding to different convective states under the same external conditions: different convective roll directions give rise to different ice front shapes (Fig.~\ref{fig3}(a) panels \text{\uppercase\expandafter{\romannumeral2}}-\text{\uppercase\expandafter{\romannumeral4}} vs \text{\uppercase\expandafter{\romannumeral3}}-\text{\uppercase\expandafter{\romannumeral1}}, and panels \text{\uppercase\expandafter{\romannumeral2}}-\text{\uppercase\expandafter{\romannumeral5}} vs \text{\uppercase\expandafter{\romannumeral3}}-\text{\uppercase\expandafter{\romannumeral2}}). This multistability (or bifurcation) of course already exists in the standard RB case where the flow can rotate either in one direction or in the opposite depending on initial tiny perturbations. When coupled with phase change, as the ice front reaches one of the shape forms, the flow is locked to fit in with this specific shape, leading to a preferred direction of convective rolls. We have not observed large-scale circulation reversals 
in the system. In the current experiments, we have no control over the convective pattern, e.g., we do not have a system to heat up the liquid locally. While in the simulations we can achieve the multiple equilibrium states by properly preparing the initial condition of the simulation. 
This interesting feature might be exploited in future experiments for flow control applications.
On the other hand, when $\Gamma<1$ the system equilibrates at smaller $H_i$. In these geometrically confined cases ($\Gamma = 0.25$ and 0.5), several convective rolls stack upon each other penetrate the stably-stratified layer and finally affect the ice front (Fig.~\ref{fig3}(a) panels \text{\uppercase\expandafter{\romannumeral2}}-\text{\uppercase\expandafter{\romannumeral1}} and \text{\uppercase\expandafter{\romannumeral2}}-\text{\uppercase\expandafter{\romannumeral2}}). 
Note that the $H_i$ from the simulations and the experiments agree well with each other except for very small $\Gamma$, which is presumably because of the stronger influence from the external environment in the experiments due to the larger area of the sidewalls. 

For the VC cases (Fig.~\ref{fig3}(b)), independently of $\Gamma$, it is observed the robust ice front morphology due to the ubiquitous presence of two counter-rotating convective rolls, which is unique to the system with density anomaly working fluids \cite{wang2020ice}. One roll originates from the cold upward convective current along the ice front (visualized by the blue arrow in Fig.~\ref{fig3}(b)); the other results from the hot plumes detaching from the hot plate (visualized by the red arrow in Fig.~\ref{fig3}(b)). The competition between these two rolls creates the ice font morphology under different aspect ratios. As $\Gamma$ increases, the interplay between these two convective rolls intensifies, leading to highly mixing of the colder and warmer fluids, as corroborated by the highly uneven $T_c$ isotherm in Fig.~\ref{fig3}(b). At $\Gamma =4$, along the wide extent of the ice front, the penetration of the hot plumes impingement can even modify the local shape of the ice front. The global ice thickness first increases a bit with increasing $\Gamma$, and then at large $\Gamma$, $H_i$ has a much weaker dependence of $\Gamma$ (Fig.~\ref{fig3}(c)). 
Note that we do not perform simulations for the aspect ratio ranging of $\Gamma<0.25$ or $\Gamma>4$, the reasons are 1) for even smaller $\Gamma$, multiple modes of convective roll configuration are observed in the classical RB as a consequence of the elliptical instability \cite{zwirner2020elliptical}, and 2) for much larger $\Gamma$, the simulations are quite expensive to carry out. Future work is required to map out a more comprehensive picture of the aspect ratio dependence, with possible extensions to three-dimensional geometries.

As observed above, the VC cases present robust forms of the ice front morphology due to the presence of two counter-rotating convective rolls. We measure the ice front shape for different aspect ratio cases we superpose them in a single figure (Fig.~\ref{fig4}(a)). The ice front morphology is similar particularly at the bottom of the cell. To understand this feature, 
we use the theoretical model proposed in \cite{wang2020ice} based on the concept of developing boundary layer (BL).  
Here we only briefly sketch the model assumptions and its construction. Although phenomenologically, it puts forward the physical mechanism that dominates the form of the ice front. It can be observed that there is a recirculating region induced by the upward convective currents adjacent to the ice (with temperature ranging from $T_0$ to $T_c$, blue arrow in Fig.~\ref{fig3}(b)). This recirculating region can be separated into the thermal BL along the ice front (orange area in Fig.~\ref{fig4}(b)) and the main flow region (light red area in Fig.~\ref{fig4}(b)). When the system reaches an equilibrium, there exists a balance between the heat flux through the BL and that through the ice. A curvilinear coordinate, $S(x)$, is introduced to measure the length of the ice front starting from its boundary point at $x=0$, and $S(x) = \int_{0}^{x} \sqrt{1+(\tfrac{d (h_i(\xi))}{d\xi})^2}\, d\xi$.  
The BL thickness is $\delta_T(S)$, which is a function of $S$, measured normal to the ice front (see Fig.~\ref{fig4}(b)).
To analyze the heat flux across the ice and the thermal boundary layer, we make the following approximations. For the ice layer, the temperature gradient normal to the ice front in the ice layer can be approximately estimated by $h_i(x) / \frac{dS(x)}{x}$; for the thermal boundary layer, which is still nested inside the viscous boundary layer ($Pr \approx 11$), the heat transfer normal to the ice front in the BL is mostly conduction \cite{sun2008experimental}.
The heat flux balance in the direction normal to the ice surface can be expressed as,
$	k_i\frac{(T_0-T_t)}{h_i(x) \frac{dS(x)}{dx} }  = k_w \frac{(T_m-T_0)}{\delta_T(S(x))}$.
$\delta_T(S)$ is assumed to be $\delta_T(S) = C_1 (S  + C_2)^{1/4}$ in which there are two parameters $C_1 = c \cdot (g [1-\rho(T_m)/\rho_c]/(\nu \kappa))^{1/4}$ with $c$ the proportional constant ($c \approx 5$), and the offset $C_2$ because of non-zero boundary layer thickness at $x=0$, with $C_2 =(h_i(0)[k_w(T_m-T_0)]/[k_i(T_0-T_t)]/C_1)^{1/4}$ \cite{bejan2013convection,white2006viscous,shishkina2016momentum}. 
Fig.~\ref{fig4}(c)-(g) show the comparison of the ice front morphology from the experiments (blue shaded area), numerical simulations (green line), and the theoretical model (red line). A good qualitative agreement is reached at the bottom of the ice front, indicating that the exact form of the ice front is indeed the thermal boundary layer dominated. The ice front morphology from the numerical simulation and from the experiment is slightly different when the aspect ratio is large (Fig.~\ref{fig4}(f) and (g)), which is presumably in that, near the large extent of the ice front, the mixing behavior between the colder (with the temperature from $T_0$ to $T_c$) and the warmer (with the temperature from $T_c$ to $T_b$) convective rolls intensifies, which is attested by the highly irregular
$T_c$ isotherm in Fig.~\ref{fig3}(b)\text{\uppercase\expandafter{\romannumeral4}}-\text{\uppercase\expandafter{\romannumeral2}}, and \text{\uppercase\expandafter{\romannumeral1}}-\text{\uppercase\expandafter{\romannumeral3}}.

\textbf{Conclusion and perspectives.} -- 
We have explored the equilibria of icing water in a differentially heated cavity by means of laboratory experiments, numerical simulations, and (phenomenologically) theoretical modeling. We showed through simulations that the effect of the initial conditions, either fully liquid or fully solidified, can lead to different equilibria, a phenomenon only observed for the RB arrangement of the system in the moderate thermal stratification case (Regime-3) and not in the VC case. 
We have also introduced a simple one-dimensional model that in part relies on the known heat-flux scaling relation in the RB system without melting, which is capable of predicting the multiplicity of equilibria in the system and their occurrence depending on the initial condition of the numerical experiment.
In the VC configuration, the system reaches the same equilibrium state independently of the system's initial condition, and this is due to the intrinsic thermal instability of this configuration, which displays convection for every Rayleigh number different from zero.\\
Secondly, we have studied experimentally and numerically the dependence of the ice front morphology on the $\Gamma$ aspect-ratio in the freezing case for the RB and VC. It was shown that for $\Gamma \geq 1$ in the RB case, the global ice thickness is nearly independent of $\Gamma$. The ice front displays large wavelength modulations and presents a local flatness in between neighboring convective rolls. Multiple forms of the ice front morphology can be observed under the same external conditions: different convective roll configurations lead to different ice front shapes. Once the system reaches a specific ice front shape, the convective roll appears locked in a certain direction. This multistability behavior provides possibilities for flow control to maintain desirable flow structures or solid-fluid interfacial morphology.
For the VC case, despite the major differences in the global ice extend as compared to RB, a robust form of the ice front morphology can be observed at all $\Gamma$. Such interfacial form is due to the presence of two counter-rotating convective rolls and is dominated by a developing boundary layer in the upward direction.  

We hope that both the methodology of the present study on water melting/freezing in idealized conditions and its findings will stimulate further investigations. More complex natural and industrial phenomena are still challenging because many crucial factors should be considered, e.g.,  the effect of persistent shear flows \cite{toppaladoddi_wettlaufer_2019,couston_JFM_2021,toppaladoddi_2021} or induced by rotation \cite{ravichandran_wettlaufer_2021}, the concentration gradients of solute components, and the extremely high-pressure in deep water body environments, which are of great relevance for the appropriate modeling of geophysical and climatological processes. 

\acknowledgments This work was supported by the Natural Science Foundation of China under Grants 11988102, 91852202, and 11861131005.
\bibliographystyle{eplbib}


	%
	\newpage

{	\Large{\textbf{Supplementary Information for:\\ Equilibrium states of the ice-water front in a differentially heated rectangular cell}
	}}
	
	\shorttitle{}

	\abstract{Supplementary Information for the main article
	}
	
		\maketitle
		
\begin{figure*}[!hb]
		\centering
		\includegraphics[width=1\linewidth]{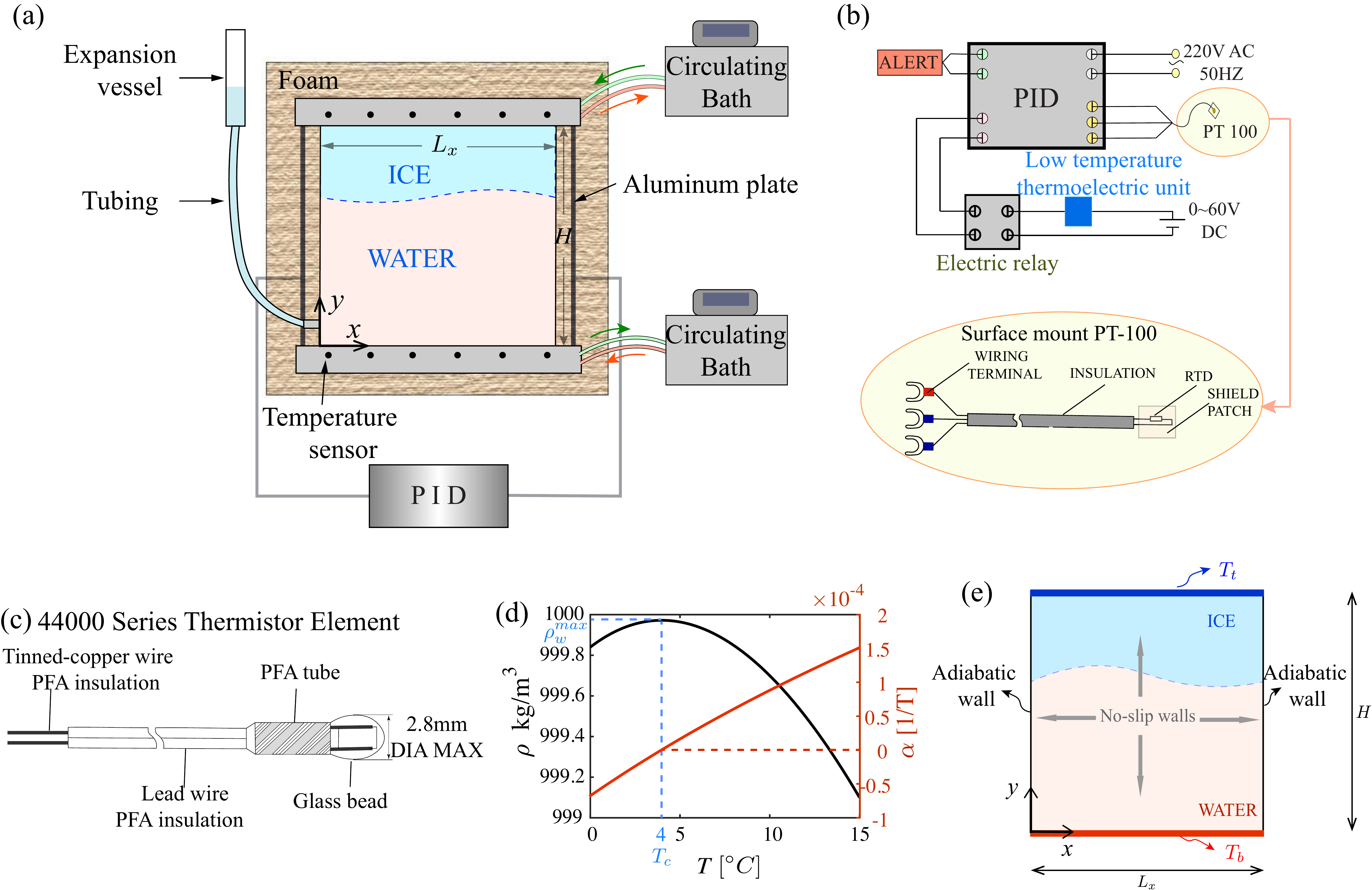}
		\caption{System configuration. (a) Sketch of the experimental system for Rayleigh-Bénard convection coupled with phase-change of pure water. 
			(b) Temperature control facilities: the PID (Proportional-Integral-Derivative) controller. (c) Sketch of resistance thermistor: 44000 series thermistor element, which is used to measure the temperature of the top and bottom plates . (d) The nonmonotonic relationship of the water density (black line) and thermal expansion coefficient (red line) with temperature for cold water near the density peak temperature, $T_c$ ($\sim 4^\circ$C at the atmospheric pressure), from Ref. \cite{Gebhart1977A}. (e) Schematic description of the problem considered in the numerical simulation (which is the same as the problem considered in the experiment). The blue shaded region corresponds to the solid phase (ice) and the red shaded region to the liquid phase (water). The top plate is cooled at constant temperature $T_t$, and the bottom plate is heated at the constant temperature $T_b$. The dashed line indicates the ice-water interface with temperature $T_0$ and is of no slip condition. The two side walls are of adiabatic conditions. The formulation for the boundary conditions read, $ T(x,0,t) = T_\text{b},~T(x,1,t) = T_\text{t},~\vec{u}(x,0,t) = 0,~\vec{u}(x,1-H_\text{i}(x,t),t) = 0,~\frac{\partial {T(x,y,t)}}{\partial x}|_{x=0} = 0,~\frac{\partial {T(x,y,t)}}{\partial x}|_{x=L_x/H} = 0,~L \rho_\text{i} V_n = \vec{n} \cdot \vec{q}_\text{w} - \vec{n}\cdot \vec{q}_{\text{i}}$, where $L$ is the latent heat, $V_n$ is the normal speed of the ice-water interface, and $h_i(x,t)$ the dimensionless ice layer thickness at the position $x$, $\vec{q}$ is the heat flux vector, $\vec{n}$ is a unit normal at the interface pointing into the liquid. The subscripts I and w refer to the ice and the water, respectively. The heat flux reads $q_\text{i} =  -k_\text{i} \vec{\nabla} T_\text{i}$ and $q_\text{w} =  -k_\text{w} \vec{\nabla} T_\text{w}$.
		}
		\label{fig1exp} 
	\end{figure*}

	\textbf{Experimental Methods.} -- The experiments for phase-change of pure water are performed in a standard Rayleigh-B\'enard (RB) convection cell (see Fig.~\ref{fig1exp}(a)-(c)).  Fig.~\ref{fig1exp}(a) shows the sketch of the experimental cell, which consists of a top cooling plate (with temperature $T_t$), a heating bottom plate (with temperature $T_b$), and plexiglass adiabatic sidewalls (four values of aspect ratio, $\Gamma=l_x/h$, are used, i.e., 0.5, 1, 2, 4 with the length, $l_x = 24$ cm, and width, $l_z = 6$ cm, which are both fixed, and the height $h = 12$ cm, $24$ cm, $48$ cm and $96$ cm, respectively, here the lowercase letter denotes physical variables with unit, while the capital letter to be mentioned later is nondimensionalized variables normalized by $h$). Gaskets are embedded in between the top plate and the sidewalls, and in between the bottom plate and sidewalls, to create an optimal seal.
	The top and bottom plates are well controlled at constant temperatures ($T_t < T_0$, and $T_b > T_0$, with $T_0$ being the phase change temperature, which is $T_0 \sim 0^\circ$C at atmospheric pressure) by the circulating bath (PolyScience PP15R-40), with the typical temperature fluctuations less than $\pm0.2$K. 
	The temperature of the top and bottom plates are monitored by the thermistors (44000 series thermistor element, see the sketch in Fig.~\ref{fig1exp}(c)). 
	The working fluid, which is confined between the top and bottom plates, is deionized and ultrapure water, of which the physical parameters (i.e., density, $\rho$, and the thermal expansion, $\alpha$) around the density peak temperature, $T_c$ ($\sim$ 4$^\circ$C at atmospheric pressure) are shown in Fig.~\ref{fig1exp}(d). The working fluid is carefully degassed by boiling twice before any experiments are performed.
	During the phase-change process, to balance the pressure induced by the volume change, an expansion vessel is connected to the experimental cell through a rubber tube, and the expansion vessel is directly open to the atmosphere so that the pressure of the experimental cell remains unchanged. 
	To reduce the heat exchange between the experimental cell and the environment, we adopt two approaches: 1) the cell is wrapped in a sandwich structure: insulation foam, aluminum plate (aiming to support and is the temperature measuring spot of the sensor, PT 100, for the Proportional-Integral-Derivative (PID) system), and insulation foam; 2) a PID controller is utilized to control the temperature of the surrounding environment outside the cell (see Fig.~\ref{fig1exp}(b)).
	Given the difficulty of the manufacture of full-cell-sized, bubble-free, and transparent ice, we did not perform the melting experiment. Instead, we mainly perform the solidification experiments of the aspect ratio dependence and investigate the bistability based on the numerical simulation and the theoretical modeling.\\

	\textbf{Numerical Methods.} -- The simulations are conducted by the \textsc{CH4-PROJECT} code \cite{calzavarini2019eulerian}, which adopts a Lattice-Boltzmann algorithm for the description of fluid and temperature dynamics, and an enthalpy method for the phase change process. The method has been validated against experiments \cite{wang2021growth, wang2020ice} and the code has been intensively tested \cite{Esfahani2018Basal}. Here, only a brief description of the method is presented and a more detailed description can be found in previous works \cite{wang2021growth, wang2020ice}. 
	The simulation is based on the Boussinesq approximation except that for the buoyancy term we use the non-monotonic relationship of the water density as a function of the temperature, $\rho = \rho_c(1-\alpha^*|T_b-4|^q )$, 
	with $\rho_c = 999.972~kg/m^3$ the maximum density corresponding to $T_c \approx 4^\circ$C, $\alpha^* = 9.30\times10^{-6} (K^{-q})$, and $q=1.895$ \cite{Gebhart1977A}. 
	The physical parameters for all simulations described in this paper are $\nu_w =1.36 \times 10^{-7}{m^2/s}, k_w=0.57 {W/(m\cdot K)}, k_i=2.37{W/(m\cdot K)}, C_{pw}=4.21 \cdot 10^{3} {J/(kg \cdot K)}, C_{pi}=1.96 \cdot 10^{3} {J/(kg \cdot K)}$.
	We neglect the microscopic physics leading to kinetic undercooling, Gibbs-Thomson effect and the anisotropic growth/melting \cite{dash2006physics} in the simulations. We implement two source terms to the energy conservation equation: 1) note that the key to accurately solve such problems is to recover the diffusion term in the energy equation exactly and, similarly to Ref. \cite{chen2017a}, the correction source term is implemented which is induced by the heterogeneous media in the investigated domain; and 2) the source term resulting from the latent heat contribution at the ice-water interface.  
	The governing equations read
	
	\begin{equation}
		\begin{split}
			&\vec{\nabla} \cdot \vec{u} = 0,\\
			& \frac{\partial \vec{u}}{\partial t}  + \vec{u}\cdot \vec{\nabla} \vec{u} = -\frac{\vec{\nabla} p}{\rho_\text{c}} +\nu_w \nabla^2 \vec{u} +\alpha^*g|T-4|^q  \hat e_z,\\
			&	\sigma (\rho \text{C}_\text{p})_0 \frac{\partial T}{\partial t} + \vec{ \nabla} \cdot (\sigma (\rho \text{C}_\text{p})_0  T\vec{u}) = \vec{\nabla} \cdot (k \vec{\nabla} T) +S_1+S_2,
		\end{split}
		\label{governing_eqn}
	\end{equation}
	where $\vec{u}(x,y,t)$, $p(x,y,t)$ and $T(x,y,t)$ are fluid velocity, pressure, and temperature fields (all temperatures are measured in Celsius), respectively and 
	we denote with $x$ as the horizontal direction and with $y$ the vertical direction; $\nu_w$, $\rho$, $g$ and $k$ are the kinematic viscosity of water, the density, the acceleration of gravity, and the thermal conductivity (when it is in water phase $k = k_{w}$, $\text{C}_\text{p}= \text{C}_\text{pw}$, and when it is in ice phase $k = k_{i}$, $\text{C}_\text{p}= \text{C}_\text{pi}$) respectively. $\hat e_z$ is the unit vector pointing in the direction opposite to that of gravity.
	The first source term is $S_1=- L\rho\frac{\partial \phi_w}{\partial t}$ and the second source term $S_2=-\sigma k \vec{\nabla} T \vec{\nabla} \frac{1}{\sigma} - \frac{\rho \text{C}_\text{p}}{\sigma} T \vec{u} \vec{\nabla} \sigma$. Here $\sigma = \frac{\rho \text{C}_\text{p}}{(\rho \text{C}_\text{p})_0}$ is the ratio of heat capacitance (which is variable and depends on the type of phase, i.e., ice or water) and $(\rho \text{C}_\text{p})_0$ is reference heat capacitance which is taken as constant \cite{chen2017a}.
	
	The boundary conditions corresponding to the governing equations above are isothermal at the top and bottom plates, no-slip at the bottom plate and at the ice-water interface, adiabatic at the lateral boundaries, and no-slip and freezing (namely, Stefan condition \cite{Alexiades1993Mathematical,bodenschatz2000recent}) at the phase-changing interface (see also Figure.~\ref{fig1exp}(e)). 
	The boundary conditions read:
	
	\begin{equation}
		\begin{split}
			& T(x,0,t) = T_\text{b},\\
			& T(x,1,t) = T_\text{t},\\
			& \vec{u}(x,0,t) = 0,\\
			& \vec{u}(x,1-h_\text{i}(x,t),t) = 0,\\
			&\frac{\partial {T(x,y,t)}}{\partial x}|_{x=0} = 0,\\
			&\frac{\partial {T(x,y,t)}}{\partial x}|_{x=L_x/H} = 0,\\
			&L \rho_\text{i} V_n = \vec{n} \cdot \vec{q}_\text{w} - \vec{n}\cdot \vec{q}_{\text{i}},
		\end{split}
		\label{governing_eqn_bcs}
	\end{equation}
	where $L$ is the latent heat, $V_n$ is the normal speed of the ice-water interface, and $h_i(x,t)$ the dimensionless ice layer thickness at the position $x$, $\vec{q}$ is the heat flux vector, $\vec{n}$ is a unit normal at the ice-water interface pointing into the liquid. The subscripts I and w refer to the ice and the water, respectively. The heat flux reads $q_\text{i} =  -k_\text{i} \vec{\nabla} T_\text{i}$ and $q_\text{w} =  -k_\text{w} \vec{\nabla} T_\text{w}$.

	We monitor the global ice thickness and the ice front morphology, which are expressed as $H_i(t)$ and $H_i(x,t)$, respectively ($H_i =1$ means full solidification). We consider that the equilibrium is reached when the standard deviation of $H_i(t)$ over a time-window of about 8 minutes is less than 0.5\%.
	
	Throughout the investigation, we limit the cold plate to a fixed temperature of $T_t = -10^\circ$C. The effects of changing $T_t$ are qualitatively predictable and are not expected to change the overall trends and mechanisms discussed in the current study. The basic effects that $T_t$ may bring are that, 1) the ice thickness at the equilibrium state of the system is thicker (thinner) with decreasing (increasing) $T_t$; and 2) the corresponding ice front morphology has a similar shape but with different extension and local curvatures. \\

\begin{figure*}[t!hb]
		\centering
		\includegraphics[width=0.5\linewidth]{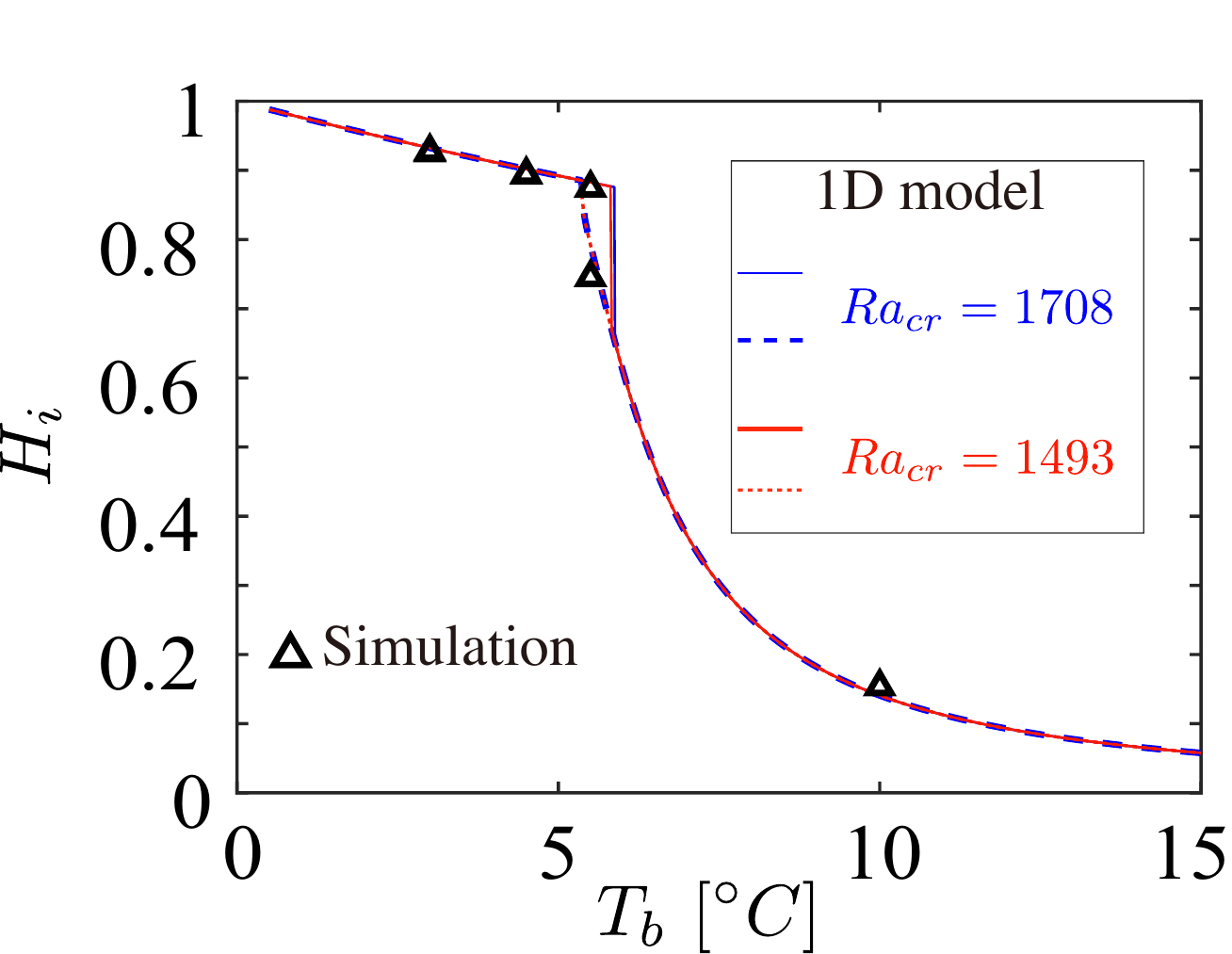}
		\caption{Theoretical prediction of the bistability phenomena for RB case: $H_i$ as a function of $T_b$ The red and blue curves correspond to the model prediction using $Ra_{cr} = 1493$ and $Ra_{cr} = 1708$ respectively, for the condition of $T_t=-10^\circ$. The symbols are results from the simulation.
		}
		\label{racr} 
	\end{figure*}

	{\color{black}
		\textbf{Comparison of one-dimensional model results based on different critical Rayleigh number} -- 
		
		In the section of Results: Bistability of the main paper, we have mentioned that the model is based on the critical Rayleigh number, $Ra_{cr} \sim1708$, which is estimated by the linear instability analysis and has been intensively validated in \cite{pellew1940maintained,dominguez1984marginal,bodenschatz2000recent}. When the system has phase change, the $Ra_{cr}$ may change in the range of 1493 to 1708 \cite{2020_Bistability}, but nevertheless it does not change the overall trend of the results in the current study.
		
		Here, we show the comparison of one-dimensional (1D) model results based on different choice of the critical Rayleigh number of $Ra_{cr} = 1493$ and $Ra_{cr} = 1708$ respectively (see Fig.~\ref{racr}). 
		It can be observed that the equilibrium state is not sensitive to the specific value of the onset of convection: with different $Ra_{cr}$ the model prediction only represents a tiny shift of the temperature range of the bistability regime. Based on this, we choose to use $Ra_{cr} = 1708$ in the calculation of the 1D model throughout the main paper.\\
	}

	\textbf{The effect of changing the cold plate temperature $T_t$} -- 
	It has been mentioned in the main paper that throughout the investigation, the cold plate is fixed at temperature $T_t = -10^\circ$C. The effects of changing $T_t$ are qualitatively predictable and are not expected to change the overall trends and mechanisms discussed in the current study. The basic effects that $T_t$ may bring are that, 1) the ice thickness at the equilibrium state of the system is thicker (thinner) with decreasing (increasing) $T_t$; and 2) the corresponding ice front morphology has a similar shape but with different extension and local curvatures. About the bistability, different $T_t$ only changes the temperature range when the bistability occurs but the overall trend remains similar.
	
	Here, by means of theoretical modeling, we explain better the effects brought by changing $T_t$.
	
\begin{figure*}[t!hb]
		\centering
		\includegraphics[width=0.7\linewidth]{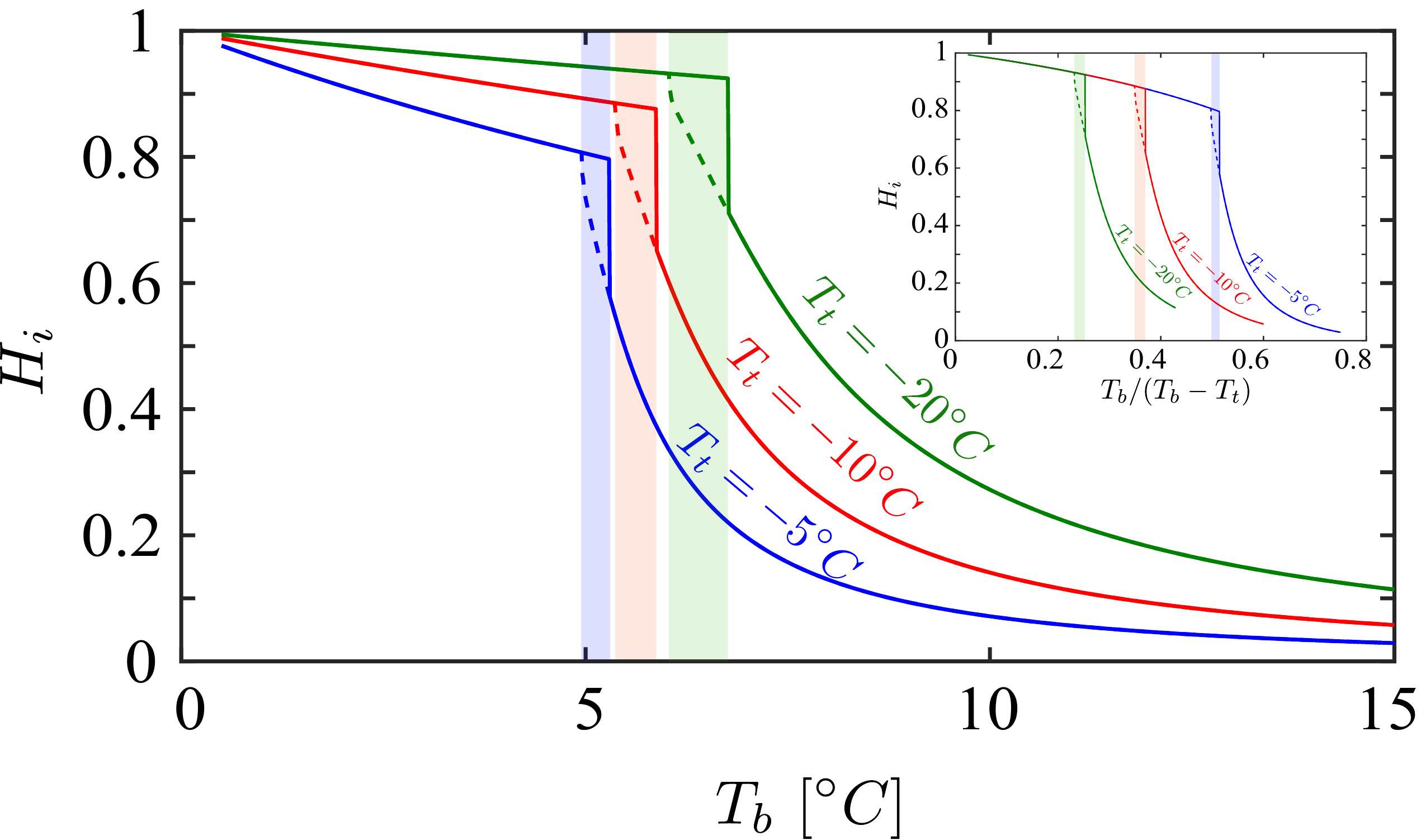}
		\caption{Theoretical prediction of the bistability phenomena for RB case: $H_i$ as a function of $T_b$ for three different cold plate temperatures, $T_t=-5^\circ$C (blue curves), $T_t=-10^\circ$C (red curves), and $T_t=-20^\circ$C (green curves). The blue, red, and green shaded areas correspond to the bisability regime for the three values of $T_t$ respectively. The inset is $H_i$ as a function of the normalized temperature $T_b/(T_b-T_t)$.
		}
		\label{changett} 
	\end{figure*}

	Fig.~\ref{changett} shows the theoretical prediction of the $H_i$ as a function of $T_b$ for three different cold plate temperatures, $T_t=-5^\circ$C (blue curves), $T_t=-10^\circ$C (red curves), and $T_t=-20^\circ$C (green curves). The blue, red, and green shaded areas correspond to the bisability regime for the three values of $T_t$ respectively. 
	
	It can be observed that a smaller amplitude of $T_t$ leads to thinner global ice thickness, and the bistability regime occurs earlier as $T_b$ increases. The temperature range of the bistability regime (shaded areas in Fig.~\ref{changett}) becomes wider as $T_t$ increases. When the $T_b$ is normalized by the temperature difference, $(T_b-T_t)$, the bistability regime has a similar range (see the inset of Fig.~\ref{changett}).

	But nevertheless, the overall trend remains similar on different $T_t$ conditions. So we do not expect the mechanisms discussed in the current study to change when varying $T_t$.\\

	\textbf{Melting and freezing in the vertical convection.} -- 
	Here, we show detailed results of the melting and freezing cases in the vertical convection.
	
	It has been mentioned in the main article that four typical cases are selected from the four distinct regimes as thermal driving increases (each regime has different levels of coupling among ice front, stably- and unstably-stratified layers \cite{wang2021growth}). To compare the Rayleigh-B\'enard convection (RB) with the vertical convection (VC), we conduct six sets of simulations for the VC case (we focus on the cases where $T_b>T_c$), with all the other external conditions remaining the same as those in the RB case, i.e., $4.5^\circ$C, $5.5^\circ$C, and $10^\circ$C, and the cold plate temperature is fixed at $T_t=-10^\circ$C. We perform long-term numerical simulations for melting (initial condition is $H_i=1$) and freezing (initial condition is $H_i=0$).
	
\begin{figure*}[t!hb]
		\centering
		\includegraphics[width=0.5\linewidth]{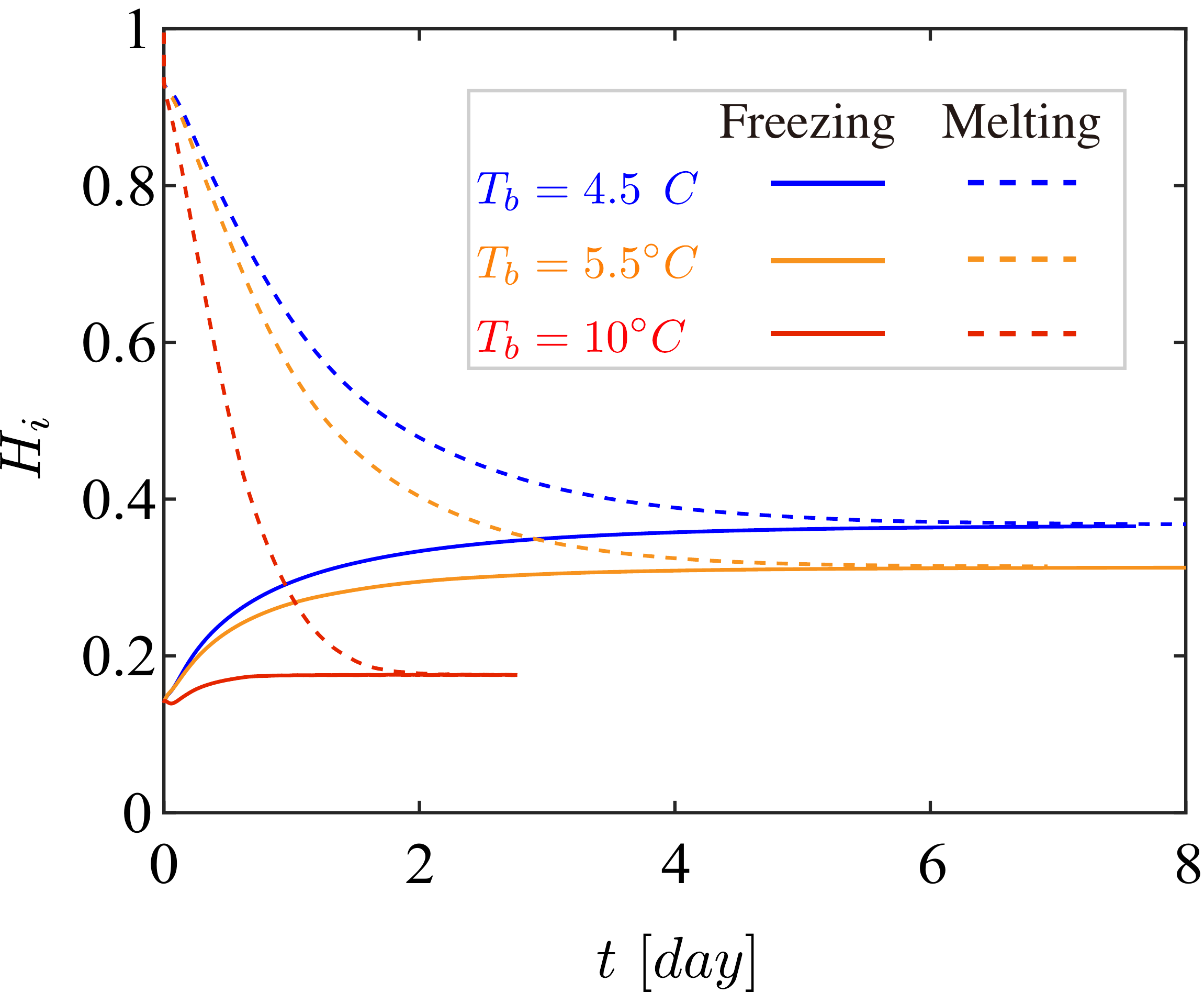}
		\caption{Evolution of the ice thickness, $H_i$, of the freezing process (thick line) and the melting process (dashed line) for the three thermal driving conditions (i.e., $4.5^\circ$C, $5.5^\circ$C, and $10^\circ$C, respectively)
		}
		\label{fig1} 
	\end{figure*}

	Fig.~\ref{fig1} shows the global ice thickness, $H_i$, as a function of time under different conditions of thermal driving in the VC case. The thick (dashed) lines represent the freezing (melting) case (denoted by ``F'' (``M'') in the legend). It is observed that the $H_i$ reaches the same final equilibrium state independent of the initial conditions or the history of the system. 
	
	We also show the evolution of the temperature field for the four thermal driving conditions (see Figs.~\ref{fig3}-\ref{fig5} for $4.5^\circ$C, $5.5^\circ$C, and $10^\circ$C, respectively). In the VC configuration, there are always two main convection rolls: 1) the colder one that is near the ice front and originates from the cold plumes detaching from the ice-water interface; 2) the warmer one originating from the hot plumes detaching from the hot plate. Different hot plate temperatures are able to adjust the relative intensity of the two convection rolls, which gives rise to different forms of ice front shapes. If the colder roll wins (Fig.~\ref{fig3}-\ref{fig4}), the ice front tends to be protected by this colder environment from deforming, and thus the ice front is flat but only tilted. On the other hand, if the warmer roll wins (Fig.~\ref{fig5}), the ice front loses the shield of the cold roll and tends to display a local maximum of the ice thickness in the middle, the thinnest at the top due to the penetration of the hot plumes, and an increasing ice thickness from the bottom.\\
	
	\textbf{Model of mechanism of bistability:} -- 
	The flow chart in Fig.~\ref{figadd} summarizes the iterative process used to compute the system equilibria. Firstly, $h_i$ varies in the range of $(0, ~h)$, and the corresponding $q_i$ can be calculated and plot as a function of $h_i$. Then an initial value of $h_s$ is set to start the iteration process. Once $h_s$ is set, $q_s$ and $q_u$ are obtained which are also a function of $h_i$. Plotting $q_s$ and $q_u$ on top of the $q_i$ vs $h_i$, if one intersection is observed, then the chosen value for $h_s$ is the solution for the equilibrium state, otherwise we update $h_s$ and begin a new round of iteration until an intersection is observed. By means of this method, it is observed two intersections when the system is in the bistability regime, while only one intersection in the convective equilibrium regime.

	\begin{figure*}[t!hb]
		\centering
		\includegraphics[width=1\linewidth]{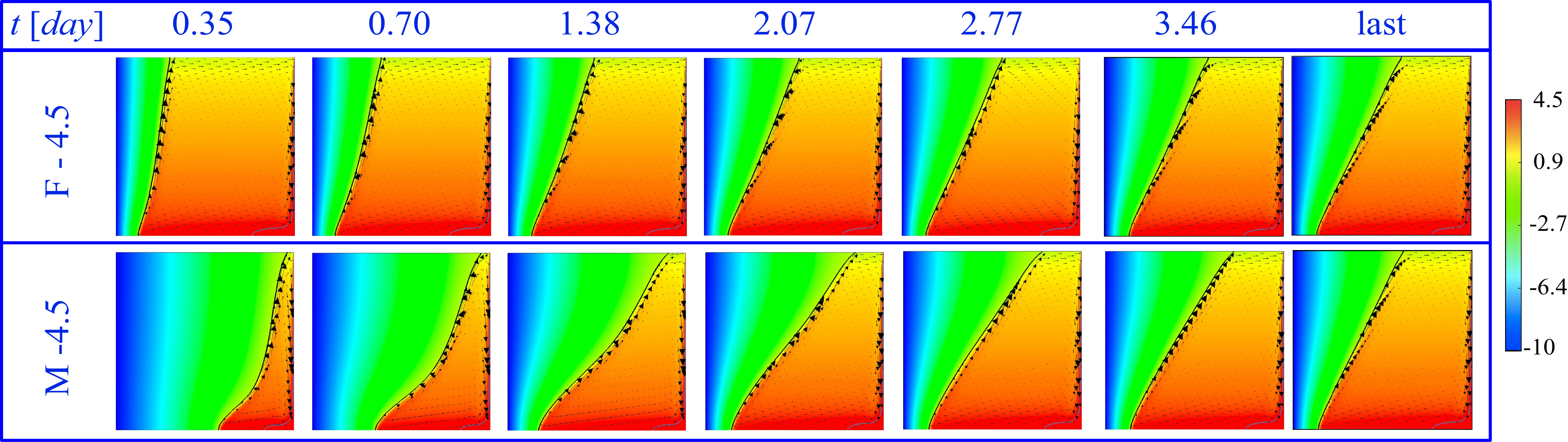}
		\caption{The evolution of the instantaneous temperature field of the freezing process (top row) and melting process (bottom row), with the bottom plate temperature $T_b = $4.5$^\circ$C. The black and blue thick lines denote the ice-water interface with temperature $0^\circ$C and the density peak temperature (i.e., $4^\circ$C) isotherm, respectively. The arrows denotes the velocity vectors.
		}
		\label{fig3} 
	\end{figure*}
	
	\begin{figure*}[t!hb]
		\centering
		\includegraphics[width=1\linewidth]{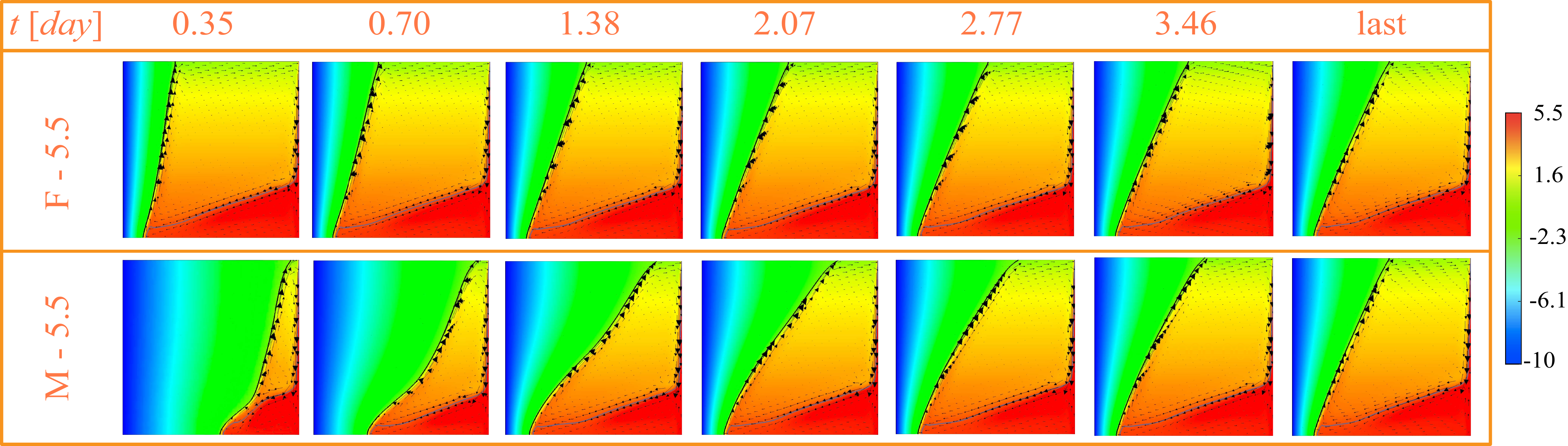}
		\caption{The evolution of the instantaneous temperature field of the freezing process (top row) and melting process (bottom row), with the bottom plate temperature $T_b = $5.5$^\circ$C. The black and blue thick lines denote the ice-water interface with temperature $0^\circ$C and the density peak temperature (i.e., $4^\circ$C) isotherm, respectively. The arrows denotes the velocity vectors.
		}
		\label{fig4} 
	\end{figure*}
	
	\begin{figure*}[t!hb]
		\centering
		\includegraphics[width=0.7\linewidth]{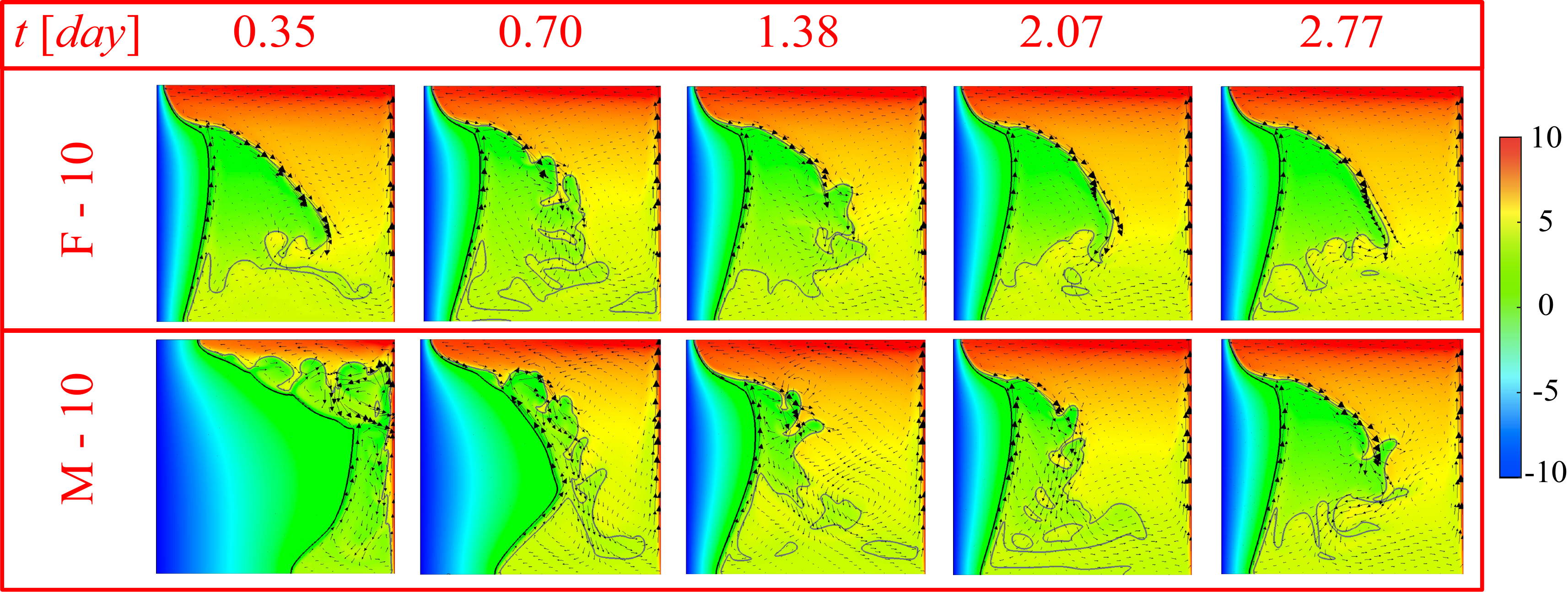}
		\caption{The evolution of the instantaneous temperature field of the freezing process (top row) and melting process (bottom row), with the bottom plate temperature $T_b = $10$^\circ$C. The black and blue thick lines denote the ice-water interface with temperature $0^\circ$C and the density peak temperature (i.e., $4^\circ$C) isotherm, respectively. The arrows denotes the velocity vectors.
		}
		\label{fig5} 
	\end{figure*}
	
	\begin{figure*}[t!hb]
		\centering
		\includegraphics[width=0.8\linewidth]{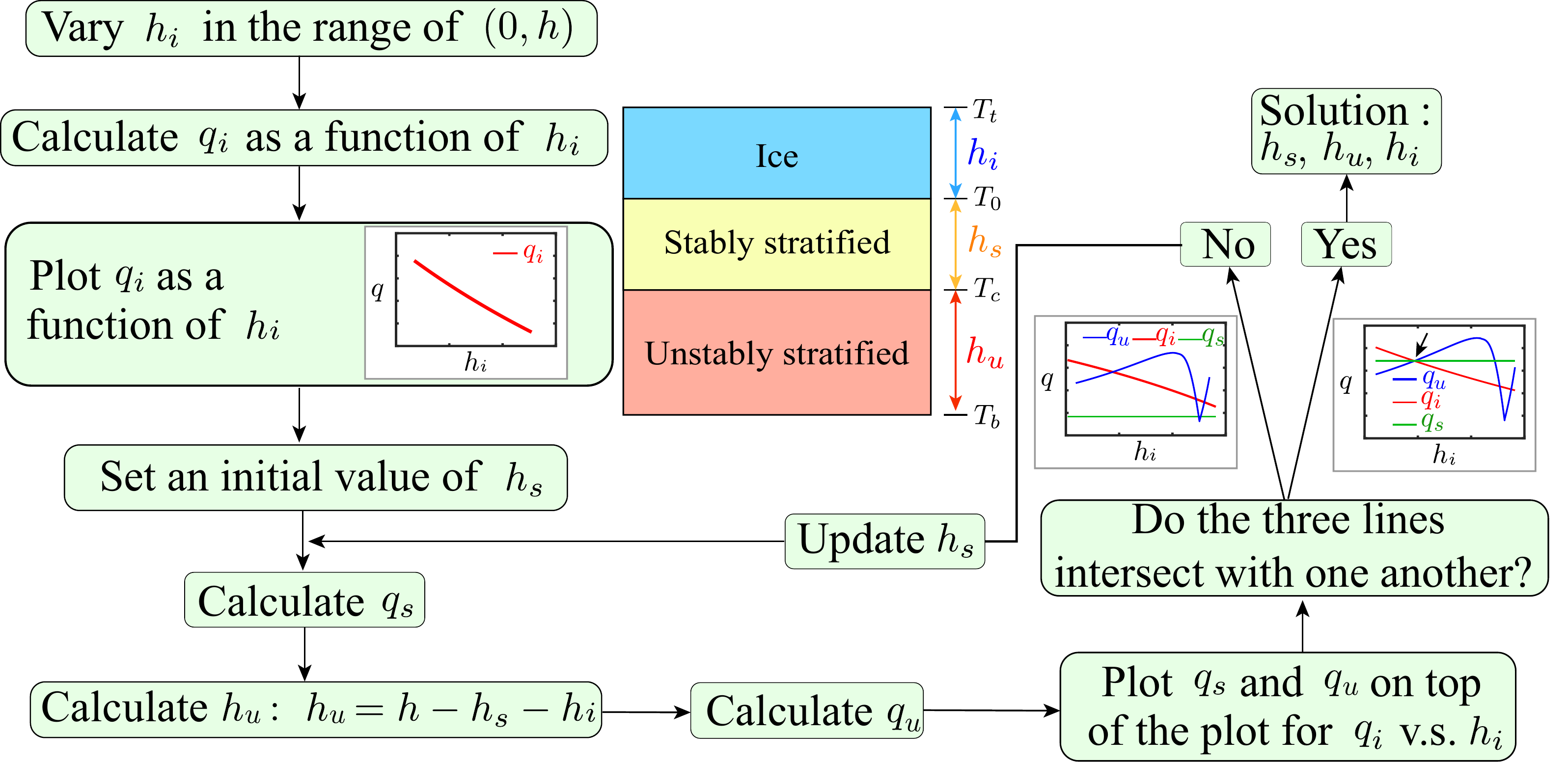}
		\caption{The flow chat that summarizes the overall iterative process to identify the solution of the thickness of the ice ($h_i$, with the temperature ranging from $T_t$ to $T_0$), of the stably stratified layer ($h_s$, with the temperature ranging from $T_0$ to $T_c$), and the unsteably stratified layer ($h_u$, with the temperature ranging from $T_c$ to $T_b$), and the corresponding heat flux through each layer is expressed as $q_i$, $q_s$, and $q_u$, respectively. This flow chart applies to the case of $T_b>T_c$ where three unknows exist (i.e., $h_i$, $h_s$, and $h_u$). When $T_b \leq T_c$, $h_c \equiv 0$, so there are only two unknowns that the heat flux $q_s$ and $q_i$ can be directly calculated without using iterations. 
			Inset: sketch of the one dimensional model: there are three layers connecting in series and we assume the interface of the neighboring layers is flat. (b) The model prediction of the conductive equilibrium that has only one intersection. 
		}
		\label{figadd} 
	\end{figure*}

	

	\bibliographystyle{eplbib}


		%
	
\end{document}